\documentclass[conference]{IEEEtran}
\usepackage{amssymb}
\usepackage{times}
\usepackage{amsmath}
\usepackage{listings}

\usepackage{graphicx}

\usepackage{url}
\urldef{\mailsa}\path|zohra.sbai@enit.rnu.tn|
\usepackage[pdfpagelabels,hypertexnames=false,breaklinks=true,bookmarksopen=true,bookmarksopenlevel=2]{hyperref}

\begin{document}

\title{Model Checking Techniques for Verification of an Encryption Scheme for Wireless Sensor Networks}
\author{
\IEEEauthorblockN{Zohra Sba\"{\i} and Mohamed Escheikh}
\IEEEauthorblockA{Ecole Nationale d'Ing\'{e}nieurs de Tunis, SysCom Laboratory\\
       BP. 37 Le Belv\'{e}d\`{e}re,
       1002 Tunis, Tunisia\\
       zohra.sbai@enit.rnu.tn\\
       mohamed.escheikh@enit.rnu.tn}
} \maketitle

 \maketitle
\thispagestyle{empty}
\begin{abstract}

In this paper, we deal with the formal verification of an encryption scheme for Wireless Sensor Networks
(WSNs). Especially, we present our first results on building a framework dedicated to modelling and
verification of WSNs aspects.
To achieve our goal, we propose to specify WSNs models written in Petri nets using Promela constructs in order to
verify correctness properties of them using SPIN Model checker.
We first specify in Promela a Petri net description of an encryption scheme for WSNs that describes its behavior. Then, correctness properties that express requirements
on the system's behavior are formulated in Linear Temporal Logic (LTL).
Finally, SPIN model checker is used to check if a specific correctness property holds for the model,
and, if not, to provide a counterexample: a computation that does not satisfy this property.
This counterexample will help to detect the source of the eventual problem and to correct it.

\end{abstract}


\section{Introduction}

Wireless Sensor Networks (WSNs) are defined as compound networks of a large number of tiny devices
called sensor nodes, which have limited processing power, storage, bandwidth, and energy \cite{RR10}.
The WSNs are widely emerging as a promising technology to stimulate
the design and the implementation of self-configuring and cost-effective
monitoring infrastructures. They are applied in many domains, such as military applications,
traffic management, ecological and environmental monitoring.
Nowadays the WSNs are increasingly being required for applications
where the data reliability needs to be guaranteed.
This can be achieved when a precise model is described and an efficient verification is ensured.
Besides, are emerging the challenges of detecting
the relevant quantities, collecting the data, assessing and evaluating the information, performing
decision making and alarm functions. Due to this, many research works
have been directed in this sense.

In this direction we focus on the description of WSNs models and their early verification.
Thus errors would be detected before execution time and hence many problems such as
accidents can be avoided.

When focus on modelling, Petri nets were originally developed to meet the need in specifying process
synchronization, asynchronous events, concurrent operations, and conflicts or resource sharing for a
variety of industrial automated systems at the discrete-event level.
For that reasons, we adopt WSNs modelled with Petri nets. As for the verification of this model,
 we can perform a checking of properties related to the behavior of the system such as 
appropriate synchronization and repetitive activities.
However there are critical design errors that are not discovered during classical testing.
That's why we refer to formal methods.
In addition, advanced wireless sensor network algorithms pose challenges to their formal modelling
and analysis, such as modelling real-time behaviors and analyzing both correctness and performance.


Formal methods and tools have been proved useful to give
high-level and precise descriptions of computer systems,
and to analyze exhaustively these systems at
early phases of the system development process.
They are a particular kind of mathematically-based techniques for the specification,
development and verification of software and hardware systems.
The use of formal methods for
software and hardware design is motivated by the expectation that, as in other engineering disciplines,
performing appropriate mathematical analysis can contribute to the reliability and robustness of a design.
They are especially promising when used in the development of
high-integrity systems where security is of great importance.

One of the well known formal methods is Model Checking.
This is a verification technique that explores all possible system states in a brute-force manner.
The principle of model checking consists on generating all possible executions of a process and checking that
the correctness specifications hold in each execution.
Thus, the main task is to determine whether the specification is satisfied by the process model.
Moreover, generating states and checking specifications can be done mechanically by a software tool.
This tool, called model checker, examines all possible system scenarios in a systematic manner.
In this way, it can be shown that a given system model truly satisfies a certain property.
There are many powerful model checkers such as NuSMV \cite{nusmv}, BLAST \cite{blast} and SPIN
\cite{H03,H97}.

SPIN is a tool for analyzing the logical consistency of concurrent systems, specifically of data
communication protocols. The system is described in a modelling language called Promela
(PROcess MEta LAnguage). This language allows for the dynamic creation of concurrent processes
which can communicate by means of channels or shared variables.
SPIN models are verified for correctness of model interactions, while minimizing the
internal functional interdependencies.
SPIN focus on the verification of asynchronous interactions of distributed software, and that's what makes it one of the most powerful tools for model checking.

Motivated by the interest of formal modelling and verification of WSNs models, we propose in this paper a model checking Technique for verification of an encryption scheme for WSNs.
We adopt a Petri net representation of the encryption scheme and specify it in Promela language.
Then, we express in LTL the correctness properties of this specification. Finally, we invoke SPIN to check if
these properties are verified by the WSN model specified in Promela.

The rest of this paper is organized as follows. Section 2 is dedicated to present preliminaries
on WSNs and encryption scheme, Petri nets, model checking and LTL. In section 3, we present an overview of Petri nets modelling of WSNs models. Section 4 describes our approach of specifying a WSN Petri net in Promela and the constructs
used to express its behavior. In section 5, we show how to formulate in LTL correctness properties
of WSNs models. Section 6 concludes the paper and gives directions to future work.

\section{Preliminaries}

\subsection{Wireless Sensor Networks and Encryption}

Wireless sensor networks (WSNs) have been drawing increasing attention due to many successful
applications \cite{JBD09, L04, SK09}.
The WSNs are defined as compound networks of a large number of tiny devices called sensor
nodes, which have limited processing power, storage, bandwidth, and energy. In addition, a
WSN might be often deployed on a large scale throughout a geographic region in hostile
environments \cite{RR10}.
According to how sensors are grouped and how the information of sensors is routed through
the network, there are two basic architectures of WSN: flat and hierarchical. In a flat architecture
all nodes have almost the same communication capabilities and resource constraints and
the information is routed by each sensor. In a hierarchical architecture, the sensor nodes are
grouped in clusters where one of the member nodes is the "cluster head". This node is responsible
for management and routing tasks.
A sensor node is composed of four basic components: sensing unit,
processing unit, transceiver unit and a power unit.

In wireless communications, encryption/decryption is very important.
Encryption is the conversion of data into a cipher-text which cannot be easily understood by unauthorized people. Decryption is the process of converting encrypted data back into its original form yielding it to be understood.
Encryption/decryption is a good idea when carrying out any kind of sensitive transaction, such as an online purchase by a credit card.

\subsection{Petri nets}

A Petri net is a 4-tuple $N=(P,T,F,W)$ where \textit{P} and \textit{T} are two finite non-empty sets of places and transitions respectively, $P\cap T = \emptyset$ , $F\subseteq (P\times T) \cup  (T\times P)$ is the flow relation, and $W : (P \times T) \cup  (T\times P) \rightarrow \mathbb{N}$ is the weight function of \textit{N} satisfying $W(x,y) = 0
\Leftrightarrow (x,y) \notin F$.

If $W(u)=1$ $\forall u\in F$ then \textit{N} is said to be ordinary net and it is denoted by $N = (P,T,F)$.

For all $x\in P\cup T$, the preset of \textit{x} is $^{\bullet}x=\{y|(y,x)\in F\}$
and the postset of \textit{x} is $x^{\bullet} =\{y|(x,y)\in F\}$.

A marking of a Petri net $N$ is a function $M:P\rightarrow \mathbb{N}$. The initial marking of $N$ is denoted by $M_{0}$.

A transition $t\in T$ is enabled in a marking $M$ (denoted by $M[t\rangle)$ if and only if
$\forall p \in$ $^{\bullet}t: M(p) \geq W(p,t).$

If transition t is enabled in marking $M$, it can be fired, leading to a new marking $M'$ such that:
$\forall p \in P: M'(p)=M(p)-W(p,t)+W(t,p)$.

The firing is denoted by $M[t\rangle M'$.

The set of all markings reachable from a marking $M$ is denoted by $[M\rangle $.

For a place \textit{p} of \textit{P}, we denote by $M_{p}$ the marking given by $M_{p}(p)=1$ and
$M_{p}(p')=0$ $\forall p'\neq p$.

Petri nets are represented as follows: places are represented by circles, transitions by boxes, the flow relation is
represented by drawing an arc between \textit{x} and \textit{y} whenever $(x,y)$ is in the relation, and the weight function
labels the arcs whenever their weights are greater than 1. A marking \textit{M} of a Petri net is represented by drawing
$M(p)$ black tokens into the circle representing the place \textit{p}.

\subsection{Model checking Techniques}

Model checking is a verification technique that explores all possible system states in a brute-force
manner. Similar to a computer chess program that checks possible moves, a model checker, the software
tool that performs the model checking, examines all possible system scenarios in a systematic manner.
In this way, it can be shown that a given system model truly satisfies a certain property.
It is a real challenge to examine the largest possible state spaces that can be treated with current
means, i.e., processors and memories. 

There are many powerful model checkers such as NuSMV, BLAST and SPIN.

SPIN model checker is a system that can verify models of computerized systems. The name SPIN was originally chosen as
an acronym for Simple Promela INterpreter. It can be used in two basic modes: as a simulator and as a verifier.
In simulation mode, SPIN can be used to get a quick impression of the types of behavior that are captured by a system model, as it is being built. Some optimization techniques, e.g., partial order reduction and graph encoding, are available to help reduce the usage of CPU time or memory space.

\subsection{Linear Temporal Logic}

\begin{figure}[h!]
\centering
\includegraphics[scale=0.8]{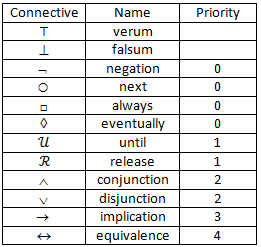}
\caption{LTL Connectives and Temporal Operators}
\label{fig:ltl}
\end{figure}

SPIN checks properties formulated in Liner Temporal Logic.
An LTL formula $f$ may contain any lowercase propositional symbol $p$,
combined with unary or binary, boolean and/or temporal operators.
LTL is built up from a set of propositional variables $p_{1}, p_{2}, ...$, the usual logic connectives
($\neg$, $\vee$, $\wedge$, $\rightarrow $ and $\leftrightarrow$) and the temporal modal operators ($\lozenge $:
eventually, $\square$: always, $\circ$: next and $\mathfrak{U}$:until) (Fig. \ref{fig:ltl}).

The semantic of a formula is given in terms of computations and the states of a computation. The atomic
propositions of temporal logic can be evaluated in a single state independently of a computation.

\section{Petri nets modelling of WSNs}


In our prospective concerning the establishment of a general modelling and verification framework of WSNs,
we propose to adopt Petri nets representations of WSNs.
In fact, Petri nets have been used for their formal semantics, graphical nature, expressiveness,
analysis techniques and tools.

In the literature, several works on the modelling of many aspects related to WSNs using Petri nets were investigated. A survey and a generalization of these works are planned to be one of the objectives of our future work.
In this paper, we propose a model checking technique for the verification of an existing Petri net model of an encryption scheme for WSNs \cite{RR10}.

\begin{figure*}[!]
\centering
\includegraphics[scale=0.6]{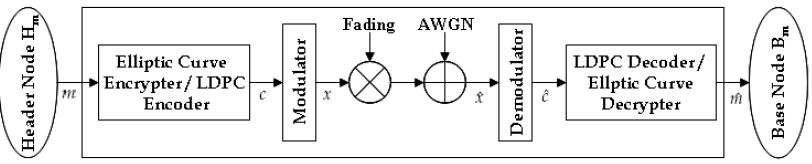}
\caption{Communication model}
\label{fig:commodel}
\end{figure*}

The adopted example (Fig. \ref{fig:commodel}) describes the secure transmission of a binary message sequence between a header node and the base station in a hierarchical WSN.
The header node (transmitter of the message) is on the left side and the base node, (receiver of the message) is on the right side.
The message sequence generated by an elliptic curve encrypter/LDPC Encoder yields to a codeword as output $c$. This codeword is converted via a modulator block in a modulator signal $x$.
Additional wireless perturbations are injected into the transmission channel model. The superposition of the signal $x$ and the fading yields to a corrupted signal $\hat{x}$. Whenever this message is received by base node, it is converted again to a binary sequence through a demodulator block called $\hat{c}$ estimating the codeword $c$. The LDPC Decoder/Ellitic Curve Decrypter uses an inverse function to transfer the binary sequence into a message sequence $\hat{m}$ providing hence an estimate of the original message $m$. The secure transmission process ends whenever the base station receives $\hat{m}$.

\begin{figure}[h!]
\centering
\includegraphics[scale=0.6]{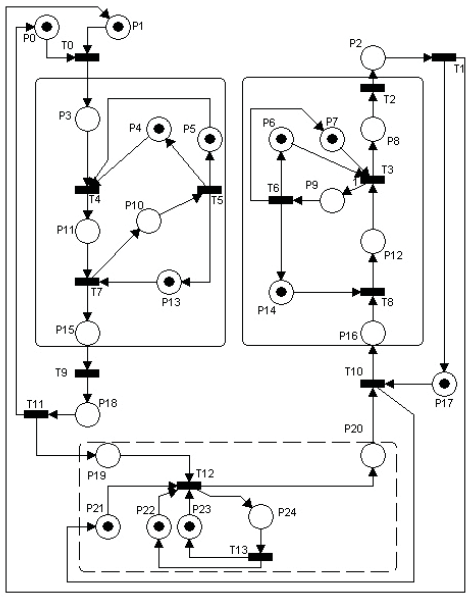}
\caption{Petri net of the communication system}
\label{fig:pncommodel}
\end{figure}

The Petri net of this secure transmission process is shown in Fig. \ref{fig:pncommodel}.
This Petri net shows the different status of the message transmitted through the communication
process. The transitions represent the tasks of transforming the message into different phases
of the process. The meanings of each place and transition are given respectively in tables \ref{pldesc} and  \ref{trdesc}.

\begin{table}[h!]
\begin{center}
\caption{The Petri net places description}
\label{pldesc}
\begin{tabular}{|c|l|c|l|}
\hline
\scriptsize{Place} & \scriptsize{Description} & \scriptsize{Place} & \scriptsize{Description}\\
\hline
P0 & \scriptsize{New message} & P13 & \scriptsize{LDPC encoder (sender side)}\\
\hline
P1 & \scriptsize{Transmitter idle} & P14 & \scriptsize{LDPC decoder (receiver side)}\\
\hline
P2 & \scriptsize{Received message} & P15 & \scriptsize{Encoder message (codeword)}\\
\hline
P3 & \scriptsize{Incoming message} & P16 & \scriptsize{Demodulated message}\\
\hline
P4 & \scriptsize{EC on the sender side (resource)} & P17 & \scriptsize{Received idle}\\
\hline
P5 & \scriptsize{Valid key (receiver public key)} & P18 & \scriptsize{Modulated message}\\
\hline
P6 & \scriptsize{EC on the receiver side (resource)} & P19 & \scriptsize{Channel input}\\
\hline
P7 & \scriptsize{Valid key (receiver private key)} & P20 & \scriptsize{Channel output}\\
\hline
P8 & \scriptsize{Decrypted message} & P21 & \scriptsize{Channel idle}\\
\hline
P9 & \scriptsize{LDPC decoder and EC restoring (receiver)} & P22 & \scriptsize{Fading process}\\
\hline
P10 & \scriptsize{LDPC decoder and EC restoring (sender)} & P23 & \scriptsize{Noise process}\\
\hline
P11 & \scriptsize{Encrypted message} & P24 & \scriptsize{Channel parameters restoring}\\
\hline
P12 & \scriptsize{Decoded message} & & \\
\hline

\end{tabular}
\end{center}
\end{table}

\begin{table}[h!]
\begin{center}
\caption{The Petri net transitions description}
\label{trdesc}
\begin{tabular}{|c|l|c|l|}
\hline
\footnotesize{Transition} & \footnotesize{Description} & \footnotesize{Transition} & \footnotesize{Description}\\
\hline
T0 & \footnotesize{Transmit message} & T7 & \footnotesize{Encode message}\\
\hline
T1 & \footnotesize{Wait for message} & T8 & \footnotesize{Decode message}\\
\hline
T2 & \footnotesize{Deliver message} & T9 & \footnotesize{Modulate message}\\
\hline
T3 & \footnotesize{Decrypt message} & T10 & \footnotesize{Demodulated message}\\
\hline
T4 & \footnotesize{Encrypt message} & T11 & \footnotesize{Send message to channel}\\
\hline
T5 & \footnotesize{Restore parameters} & T12 & \footnotesize{Perturb Message}\\
\hline
T6 & \footnotesize{Restore parameters} & T13 & \footnotesize{Reset channel parameters}\\
\hline
\end{tabular}
\end{center}
\end{table}

In the next section, we present how to specify in Promela a WSN modelled as a Petri net.

\section{WSN model based on Promela}


The modelling of Petri nets with Promela language is first introduced by Holzmann in \cite{H03, H97}.
In this work, a Petri net is represented as a single process describing each firing of its transitions.



In \cite{YYT08}, the authors rewrite the Petri net representation given in \cite{H03, H97} by representing the places set $P$ by an array of $|P|$ elements rather than $|P|$ byte variables.
The Promela specification of a Petri net given in \cite{H03} describes only the marking of places and
does not take into account the firing count of transitions. However, it is necessary to save the firing count of each transition while progressing in the execution of the process.
Such information will be used in the verification of the correctness properties.

We now present our Promela specification of the Petri net's description of the encryption scheme presented in the previous section. This specification consists on an application of our general approach of mapping from Petri nets to Promela presented in \cite{Z10}.

We describe (in Promela) a Petri net in terms of the marking of places and the firing count of transitions. For that, we represent the places (resp. the transitions) by an array $PL$ (resp. $TR$) of integers with length equal to the number of places (resp. transitions). These arrays contain initially zero in each element.

The behaviour of a Petri net can be described in terms of system states and their changes. These changes influence the elements of tables $PL$ and $TR$ (initialized to zero).
A marking is initialized to $M_{i}$ and it is changed according to the following transition rule:
A transition $t$ is enabled if each input place $p$ of $t$ is marked. Moreover, an enabled transition
$t$ may fire or not, and a firing of $t$ removes one token from all $p\in$ $^{\bullet}t$ and adds one
token to each output place of $t$.

These concepts are translated in the corresponding Promela description of a Petri net as follows:
The firing of a transition $t$ 
 consists on decreasing by $1$ the integer
corresponding to each place
$p\in$ $^{\bullet}t$  in the array $PL$ and increasing by $1$ the elements of $PL$ corresponding to the output places
($p\in t^{\bullet}$).
Although, we increase by $1$ the element of $TR$ corresponding to the transition $t$ to mark that $t$ is fired once.

These modifications are ensured by the macro definitions $fire$, $add1$, $add2$, ..,$addS$, $remove1$, $remove2$, .., and $removeK$ where
$S$ is the maximum number of possible input places and $K$ is the maximum number of possible output places.

If a transition $t$ has $I$ input places and $J$ output places, the firing of $t$ is ensured by a call to the following macros with the
appropriate arguments:
\begin{enumerate}
  \item $removeI(p_{1}, p_{2}, .., p_{I})$ destructs one token from each input place.
    This destruction is allowed only if these input places are marked ($PL[index~of~p_{j}]>0$, $1\leq j\leq I$).
  \item $fire(t)$ increases by $1$ the element corresponding to $t$ in $TR$.
  \item $addJ(p_{1}, p_{2}, .., p_{J}$) produces one token in each output place.
\end{enumerate}

Note that the first action contains the firing condition, thus the second and third actions are executed only if the first one is achieved.

To map the skeleton of any Promela Petri net specification, we propose to draw it in BNF (Backus
Naur Form) in which $<Process>$ is the axiom (initial symbol):

\begin{center}
\begin{tabular}{lll}
\hline

 \lstset{basicstyle=\ttfamily, basicstyle=\footnotesize,language=Promela}
\begin{lstlisting}
<Process> ::= init {
              <Initial_Marking>
              do
                <Firings>
              od
	      }
<Firings> ::= <Firing> <Firings>
<Firing>  ::= :: atomic {
                 <Enabled_Tk> -> <Fire_Tk>
                 }
\end{lstlisting}
\\
\hline

\end{tabular}
\end{center}

The $init$ keyword is used to declare the
behavior of a process that is active in the initial system state.
In this process, we propose to describe the firings by a $do$
loop in which each line specifies one transition firing. Hence, Each line describes the statements
performed to ensure the actual transition firing.
We prefix the sequence of statements ensuring the firing by the Promela construct $atomic$ in order to guarantee that the sequence of composed actions has to be executed as one indivisible unit,
non-interleaved with any other processes.

\begin{center}

\begin{figure*}[!]
\centering
\includegraphics[scale=0.7]{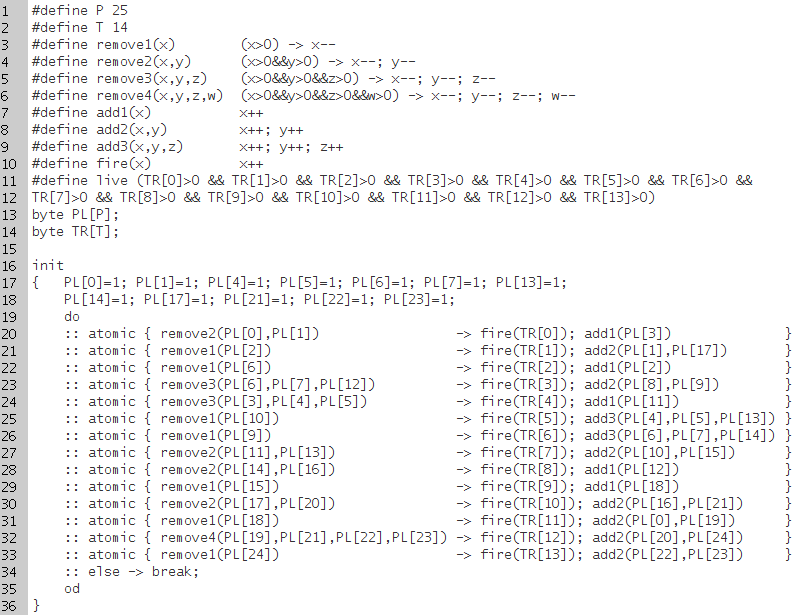}
\caption{Promela model relative to Petri net of Fig. \ref{fig:pncommodel}}
\label{fig:promela1}
\end{figure*}

\end{center}

To illustrate this Promela description of a Petri net and its behavior, we present in Fig.
\ref{fig:promela1} the Promela description of the Petri net in Fig. \ref{fig:pncommodel}.

Once presenting how to specify the encryption scheme for WSNs in Promela, we expose in the next section
the LTL formulation of correctness properties.

\section{LTL based WSN verification}



Properties to be verified by SPIN have to be expressed in Linear Temporal Logic. Once a LTL formula is specified, SPIN provides an algorithm for converting it into equivalent finite-state automata over infinite words (B\"{u}chi automata). Never claim is the Promela model of the B\"{u}chi
automata corresponding to the LTL formula which is used to specify system behavior which should never occur. When generating a verifier, SPIN creates one active process for the never claim declared. The never claim process will be executed by the verifier between every execution of other processes yielding the verifier to report error if the never claim process ends.

In concrete terms, suppose we want to express the property $Prop$ which states that the
place $P23$ of the Petri net given in Fig. \ref{fig:pncommodel}
will eventually always be marked. Then, a proposition $p$ which specifies that place $P23$ is marked should be written in Promela as follows:

\begin{center}
$\#define~p~(PL[23]~>=~1)$\\
\end{center}

\begin{figure}[h!]
\centering
\includegraphics[scale=0.3]{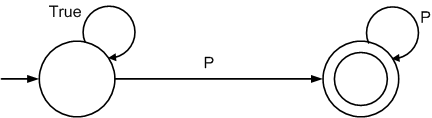}
\caption{A b\"{u}chi automata corresponding to formula F}
\label{fig:buchip}
\end{figure}

The LTL formula of $Prop$ is $F$: $<>[~]p$.
It corresponds to the b\"{u}chi automata of Fig. \ref{fig:buchip}.

A never claim consists on a negation of the property, which is in our case the following LTL formula:

\begin{center}
$!<>[~]p=[~]![~]p=[~]<>!p$\\
\end{center}

\begin{figure}[h!]
\centering
\includegraphics[scale=0.3]{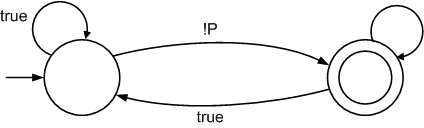}
\caption{A b\"{u}chi automata corresponding to the negation of F}
\label{fig:buchinotp}
\end{figure}

This formula corresponds to the b\"{u}chi automata of Fig. ~\ref{fig:buchinotp}.

\begin{figure}[h!]
\centering
\includegraphics[scale=0.8]{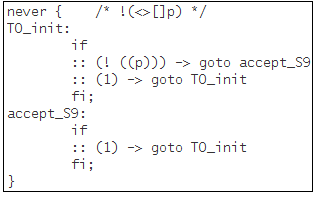}
\caption{The never claim corresponding to F} \label{fig:neverclaimp}
\end{figure}

The never claim (generated by Spin) corresponding to formula $F$ is
given in Fig. \ref{fig:neverclaimp}.

~~\\

Now referring to the proposed Promela specification, added by propositions description by means of macro
definition $\#define$, we can formulate in LTL different kinds of properties. Such properties include
but are not limited to liveness (deadlock freedom) and safeness.

\subsection{Verification of deadlock freeness property}

The deadlock freeness property states that it should be possible to execute a random task by following
the appropriate route through the Petri net i.e. there are no dead transitions. In other terms, we have to verify that each transition will fire at least once (i.e. in an execution path).
So, we may check if the following formula is verified:

\begin{center}
\fbox{$\lozenge ~live$}
\end{center}

where $live$ is a proposition defined in Promela as follows:

\begin{center}
\fbox{ $\#define~live~(\overset{j=|T|-1}{\underset{j=0}{\&\&}}  \;
(TR[j]>0))$ }
\end{center}

\begin{figure}[h!]
\centering
\includegraphics[scale=0.58]{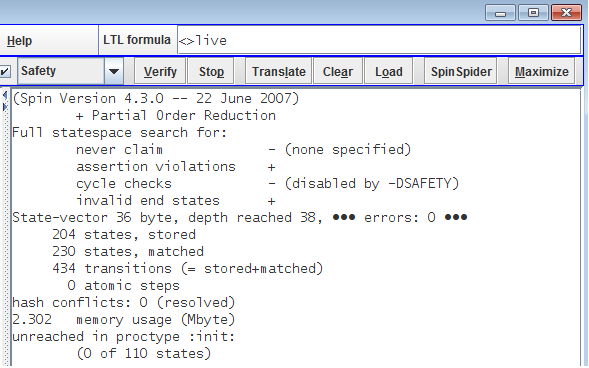}
\caption{Verification of the deadlock freeness of Petri net of figure \ref{fig:pncommodel}}
\label{fig:livecommodel}
\end{figure}

The verification of the deadlock freeness property for the Petri net of Fig \ref{fig:pncommodel} is given
in Fig.   \ref{fig:livecommodel}. In this figure, showing a part of our example verification by SPIN,
we see that the corresponding Petri net is live. This is mentioned by: $\bullet\bullet\bullet~errors:~0~\bullet\bullet\bullet$.

\subsection{Safeness property}

The property of safeness can be determined for both individual places and the entire net. A place
is said to be safe if for all possible markings the number of tokens in that place never exceeds one.
The Petri net is said safe if all the places in the net are safe.

In LTL, we express safeness by the following formula:

\begin{center}
\fbox{$\square ~safe$}
\end{center}

where $safe$ is a proposition defined in Promela as follows:

\begin{center}
\fbox{ $\#define~safe~(\overset{i=|P|-1}{\underset{i=0}{\&\&}}  \;
(PL[i]<=1))$ }
\end{center}

For the overall Petri net (Fig. \ref{fig:pncommodel}) the safeness is verified.

\section{Conclusion}

Wireless Sensor Networks are increasingly being required for applications where security and data
integrity need to be guaranteed. Hence, the formal verification of these important properties
for WSN models is promising since it can avoid critical problems by detecting them at an early stage
yielding to their correction before execution stage.

In this context, we presented in this paper an approach for the verification of an encryption scheme
for Wireless Sensor Networks modelled by Petri nets using one of the main powerful formal methods: model checking.
More precisely, we have shown that model checking is a powerful technique to verify and to detect the
eventual errors at an early stage and that a simple study of the generated counterexample permits to
correct these errors before the system deployment.

We first have shown how to write a Promela description of a given WSN model specified by a Petri net.
Second, we have studied the properties of deadlock-freeness and safeness of the studied WSN model and expressed these properties in Linear Temporal Logic.

Future work will include verification of other properties related to WSNs such as integrity
and security properties.

\end{document}